# Effect of Rb and Cs-doping on superconducting properties of $MgB_2$ thin films

**R K Singh, Y Shen, R Gandikota, D Wright, C Carvalho, J M Rowell and N Newman**

School of Materials, Arizona State University, Tempe, Arizona 85287-8706, USA

E-mail: Nathan.Newman@asu.edu

**Abstract.** Our Rutherford backscattering spectrometry (RBS) study has found that concentrations up to 7 atomic % of Rb and Cs can be introduced to a depth of ~700 Å in $MgB_2$ thin films by annealing in quartz ampoules containing elemental alkali metals at <350 °C. No significant change in transition temperature ($T_c$) was observed, in contrast to an earlier report of very high $T_c$ (>50 K) for similar experiments on $MgB_2$ powders. The lack of a significant change in $T_c$ and intra-granular carrier scattering suggests that Rb and Cs diffuse into the film, but do not enter the grains. Instead, the observed changes in the electrical properties, including a significant drop in $J_c$ and an increase in $\Delta\rho$ ($\rho_{300}$-$\rho_{40}$), arise from a decrease in inter-granular connectivity due to segregation of the heavy alkaline metals and other impurities (i.e. C and O) introduced into the grain boundary regions during the anneals.

## 1. Introduction

A recent report [1] of very high (>50 K) onsets of superconductivity in Rb and Cs-doped $MgB_2$ has caused other researchers to study the effect of heavier alkali metals on $MgB_2$ superconducting properties. According to Palnichenko et al. [1], the very electropositive heavier alkali and alkaline earth metals can donate carriers to the electron system, thereby enhancing the superconducting properties of the host $MgB_2$ material. A significantly enhanced $T_c$ was reported in bulk samples by treating $MgB_2$ powder with Rb, Cs and Ba through a liquid phase reaction. The concentration and spatial distribution of the dopants were not reported in that work.

In the study reported here, we have studied the effect of Cs and Rb on $MgB_2$ electrical properties by diffusing them into thin films grown by MBE (Molecular Beam Epitaxy). A quantitative determination of the dopant concentration has also been made using Rutherford backscattering spectrometry (RBS).

## 2. Experiment

$MgB_2$ films of thickness 1400 to 2200 Å were deposited on (0001) sapphire at either room temperature (~25 °C) or 300 ± 2 °C in an ultra-high vacuum MBE system with an ultimate base pressure of ~5×10$^{-10}$

Torr. The system pressure reached as high as $10^{-6}$ Torr during deposition. Further growth details have been published elsewhere [2,3]. The thickness and the depth profile of the chemical composition were determined using Rutherford backscattering spectrometry. Nuclear resonant elastic scattering, $^{16}O(\alpha,\alpha)^{16}O$ and $^{12}C(\alpha,\alpha)^{12}C$ was used for oxygen and carbon detection respectively. This technique can detect up to 0.5% of impurity with accuracy greater than 95%. The aerial atomic density is directly determined by RBS and then a thickness can be inferred using the known material density. Some films were patterned into a 100 μm x 1 mm bridge using photolithography and reactive ion etching (argon plasma) to provide for four-point contact measurements of resistivity and critical current density ($J_c$).

In the inert atmosphere glove box, these films were loaded into rigorously clean clear fused quartz ampoules with an elemental Mg chip and approximately 0.5 cc of alkali metal (Rb or Cs). The films were kept away from the metals by means of individual quartz containers within the ampoule separated by a quartz spacer. The ampoule with contents was then transferred to a high vacuum manifold and evacuated to $5\times10^{-5}$ Torr and sealed using a hydrogen/oxygen torch. A quartz handle was attached to allow for control of the ampoule while in the vertical 3-zone clamshell tube furnace used to anneal the ampoule and during quenching. Annealing was done over a wide temperature range (100 – 350 $^o$C) and time (12 – 100 hours). At the time of quenching, the metal end of the ampoule was lowered into ice water before the region containing the thin film such that metal vapour condensed on the colder end of the ampoule instead of the thin film. Four-terminal current-voltage measurements were made using either a Quantum Design physical property measurement system (PPMS) or a custom-built dipping probe using direct contacts in un-patterned films and silver contacts in patterned films.

## 3. Results and discussion

*3.1 Chemical composition of annealed films*

Table 1 summarizes the annealing treatment and post-annealed composition of the $MgB_2$ films. One sample (X-1) was annealed without any dopant to distinguish the contribution of the post-growth thermal treatments from doping-induced changes in the $MgB_2$ superconducting properties. We observe an inhomogeneous distribution of Rb or Cs with depth in all the samples studied. The greatest concentration of Rb-doping (7 atomic % in top 700 Å layer and 4.3 atomic % in bottom 700 Å layer) was achieved in the film Rb-4, after a thermal treatment of 350 $^o$C for 15 hours followed by water quenching the ampoule. When a similar thermal treatment was given to $MgB_2$ film grown at room temperature (Rb-3), considerably lower (<1 atomic %) Rb was observed in the film. In the case of Cs, a maximum surface doping (4.7 atomic %) and sub-surface doping (0.8 atomic %) was achieved after thermal treatments at 160 $^o$C for 20 hours followed by 100 $^o$C for 100 hours (Cs-3) and 200 $^o$C for 18 hours (Cs-1) respectively. In some samples, we also observe changes in film thickness after thermal treatment. This increase in thickness is presumably a result of the reaction of the film with the ambient atmosphere or the ampoule. However, the reason for the reduction in the aerial density and the inferred thickness in the sample annealed at 300 $^{\circ}$C (Cs-2) is not clear. The vapour pressure of thin films of $MgB_2$ at this temperature has been shown to be very small (<<$10^{-3}$ monolayer per second) [4].

**Table 1.** Chemical composition of $MgB_2$ films before and after thermal treatment.

| # | Thermal treatment | Composition & thickness before treatment | Composition & thickness after treatment |
|---|---|---|---|
| X-1[a] | 300 $^{\circ}$C/15 hours, water quenched | $Mg_{1.05}B_2O_{0.3}$ (1450 Å) | $Mg_1B_2O_{0.5}$ (1470 Å) |
| Rb-1 | 200 $^{\circ}$C/12 hours, air cooled | $Mg_1B_2O_{0.04}$ (2150 Å) | $Mg_{1.05}B_2Rb_{0.09}O_{0.2}C_{0.03}$ (top 250 Å) $Mg_{1.05}B_2Rb_{0.01}O_{0.04}$ (middle 1500 Å) $Mg_{1.05}B_2O_{0.04}$ (bottom 400 Å) |
| Rb-2 | 300 $^{\circ}$C/15 hours, water quenched + air annealed 200 $^{\circ}$C/7 hours | $Mg_1B_2O_{0.3}$ (1700 Å) | $Mg_1B_2Rb_{0.004}O_{0.3}$ (top 800 Å) $Mg_1B_2Rb_{0.007}O_{0.3}$ (bottom 900 Å) |
| Rb-3[b] | 350 $^{\circ}$C/15 hours, water quenched | $Mg_{1.4}B_2O_{0.015}$ (1700 Å) | $Mg_{1.1}B_2Rb_{0.025}O_{0.6}C_{0.2}$ (top 750 Å) $Mg_{1.1}B_2Rb_{0.035}O_{0.6}C_{0.2}$ (bottom 750 Å) |
| Rb-4 | 350 $^{\circ}$C/15 hours, water quenched | $Mg_1B_2O_{0.16}$ (1400 Å) | $Mg_1B_2Rb_{0.5}O_{2.5}C_{1.1}$ (top 700 Å) $Mg_1B_2Rb_{0.2}O_1C_{0.4}$ (bottom 700 Å) |
| Cs-1 | 200 $^{\circ}$C/18 hours, water quenched | $Mg_1B_2O_{0.015}$ (1400 Å) | $Mg_1B_2Cs_{0.12}O_1C_{0.35}$ (top 800 Å) $Mg_1B_2Cs_{0.03}O_{0.5}C_{0.2}$ (bottom 1200 Å) |
| Cs-2 | 300 $^{\circ}$C/18 hours, water quenched | $Mg_1B_2O_{0.015}$ (1400 Å) | $Mg_1B_2Cs_{0.13}O_1C_1$ (top 60 Å) $Mg_1B_2Cs_{0.0012}O_{0.05}C_{0.02}$ (bottom 1100 Å) |
| Cs-3 | 160 $^{\circ}$C/20 hours + 100 $^{\circ}$C/100 hours, water quenched | $Mg_{1.1}B_2O_{0.13}$ (3200 Å) | $Mg_{1.1}B_2Cs_{0.17}O_{0.25}C_{0.08}$ (top 300 Å) $Mg_{1.1}B_2Cs_{0.007}O_{0.15}C_{0.02}$ (next 1200 Å) $Mg_{1.1}B_2Cs_{0.001}O_{0.15}$ (bottom 2200 Å) |

[a]Sample annealed without dopant.
[b]Sample grown with substrate at room temperature.

We observe clear evidence of Cs diffusing into the substrate after all thermal treatments (160 – 300 $^{\circ}$C). However, diffusion of Rb into the substrate is seen only at temperatures above 200 $^{\circ}$C. In the Rb-doped sample annealed at 200 $^{\circ}$C (Rb-1), Rb could not be detected in the bottom 400 Å of the film. Figure 1 shows RBS analysis of the Cs-doped sample with high level of oxygen contamination in the $MgB_2$ film in conjunction with a significant amount of Cs that has diffused into the sapphire substrate.

A significant oxygen and carbon contamination is consistently observed in the films containing large Cs and Rb-doping concentration (table 1). The increase in carbon and oxygen generally scales in direct proportion to the Rb and Cs content. The highest reported doping level of 7 atomic % Rb (Rb-4) has ~30 atomic % increase in oxygen and ~15 atomic % increase in carbon after exposure to the doping process. Similarly, a film with a ~2.7 atomic % Cs doping level (Cs-1) is found to have ~22 atomic % increase in oxygen and ~8 atomic % increase in carbon after exposure to the doping process. There are, however, two exceptions to this - room temperature grown $MgB_2$ has a relatively higher level of contamination and the Cs doped sample annealed at low temperatures (160 $^{o}$C, 20 hours + 100 $^{o}$C, 100 hours) has a relatively lower contamination.

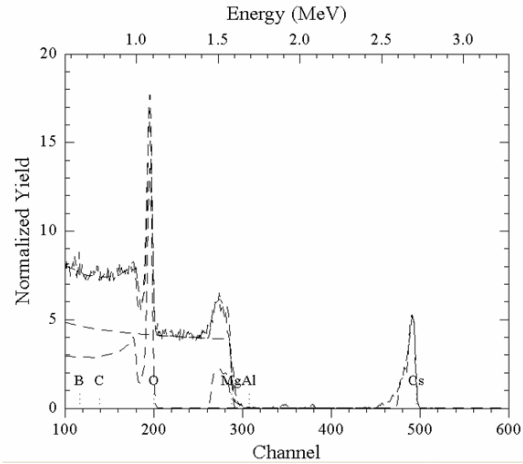

**Figure 1.** RBS data (solid) and simulation (dashed) of Cs doped sample showing (a) high oxygen (22 atomic %) in the film and (b) Cs diffusing into the substrate.

*3.2 Effect of dopants on superconducting properties*

Table 2 summarizes the effect of Rb/Cs incorporation on the critical temperature ($T_c$), resistivity ($\rho_{300K}$, $\rho_{40K}$, $\Delta\rho$, and $\rho_{40K, corrected}$), critical current density ($J_c$), and upper critical field slope ($dH_{c2}/dT$) of $MgB_2$ films. Critical current density ($J_c$) measurements were only performed on selected films that could be patterned into a 100 μm x 1 mm bridges. Properties that could not be measured as a result of the rapid degradation of the films, presumably as a result of exposure and subsequent reaction with the air, are left blank in the table.

*3.2.1 Effect on transition temperature.* For both Rb and Cs-doping, we see a small, but measurable, change (<2 K) in the transition temperature. A similar change is also observed in the sample that was annealed without dopants. It is evident, therefore, that the observed changes are a result of thermal treatment and not necessarily a direct consequence of the presence of Rb or Cs in the film.

It is interesting to note that the width of the transition appears to decrease slightly after introducing Cs (from $\Delta T_c$ = 1.4±0.2 K to 1.1±0.1 K), while it increases significantly during the thermal processing in the absence of a dopant and when Rb was introduced (to $\Delta T_c$ = 3.1±0.6 K).

*3.2.2 Effect on resistivity.* We observe a large increase (>100%) in the resistivity at 300 K ($\rho_{300}$) and 40 K ($\rho_{40}$) in all the samples, including the annealed un-doped sample. This indicates that the increased resistivity arises primarily from the thermal treatment and is not necessarily related to the presence of Rb or Cs in the film. The increase in resistivity is relatively small (20%) in the sample that was annealed with Cs at 160 °C for 20 hours followed by 100 °C for 100 hours. This is presumably a consequence of being exposed to the relatively low annealing temperatures in this case.

**Table 2.** Effect of Rb or Cs incorporation on the critical temperature, resistivity, critical current density, and upper critical field slope of $MgB_2$ thin films.

| # | Condition | $T_c$ (K) Onset | Finish | $\rho_{300K}$ ($\mu\Omega.cm$) | $\rho_{40K}$ ($\mu\Omega.cm$) | $\Delta\rho$ ($\mu\Omega.cm$) | $\rho_{40K,Corrected}$ ($\mu\Omega.cm$) | $J_c$ @ 10K (A/cm$^2$) | $dH_{c2}/dT$ (T/K) |
|---|---|---|---|---|---|---|---|---|---|
| X-1 | As-grown | 32.8 | 31.2 | 98 | 75 | 23 | 24 | - | 1.3 |
|     | Annealed | 34.5 | 30.8 | 680 | 570 | 120 | 35 | $1.6 \times 10^5$ | 1.2 |
| Rb-1 | As-grown | - | - | - | - | - | - | - | - |
|      | Annealed | 34.9 | 31.4 | 41 | 26 | 15 | 13 | - | 1.1 |
| Rb-2 | As-grown | 34.1 | 32.8 | 120 | 88 | 27 | 24 | $1.7 \times 10^6$ | 1.1 |
|      | Annealed | 35.0 | 32.6 | 250 | 230 | 20 | 84 | $6.8 \times 10^4$ | 1.2 |
| Rb-3 | As-grown | - | - | - | - | - | - | - | - |
|      | Annealed | - | - | - | - | - | - | - | - |
| Rb-4 | As-grown | 32.8 | 31.2 | 98 | 75 | 23 | 24 | $1.1 \times 10^6$ | - |
|      | Annealed | 34.5 | 30.1 | 680 | 560 | 120 | 35 | $5.8 \times 10^5$ | 1.2 |
| Cs-1 | As-grown | - | - | - | - | - | - | - | - |
|      | Annealed | 34.3 | 33.3 | 170 | 130 | 37 | 26 | - | 0.9 |
| Cs-2 | As-grown | - | - | - | - | - | - | - | - |
|      | Annealed | 34.7 | 33.7 | 300 | 250 | 58 | 31 | - | 1.0 |
| Cs-3 | As-grown | 35.4 | 33.8 | 150 | 110 | 37 | 22 | - | 0.9 |
|      | Annealed | 35.8 | 34.6 | 180 | 130 | 46 | 22 | - | 0.9 |

We also observe an increase in $\Delta\rho$ ($\rho_{300}$-$\rho_{40}$) in post-annealed films, except in Rb-2 sample, where a marginal drop in $\Delta\rho$ is observed. $\Delta\rho$, as pointed out by Rowell [5], is a measure of the inter-grain connectivity in $MgB_2$ samples. An increase in $\Delta\rho$ reflects a decrease in the connectivity, presumably as a result of oxidation or other forms of contamination segregating to the grain boundaries. If we compare changes in $\Delta\rho$ (table 2) with the extent of impurity contamination that occurs during annealing (table 1), we see that Rb-2 is the only sample where oxygen and carbon concentration did not change after annealing. This sample shows a drop in $\Delta\rho$ after annealing. In all other samples, we see a large increase in oxygen and carbon after annealing that is strongly correlated with the large increase in $\Delta\rho$. It appears, therefore, that oxygen and carbon incorporation into the film contaminates grain boundaries and thus reduces inter-grain connectivity. The change in inter-grain connectivity is found to scale proportionately with oxygen and carbon concentration. The largest change found for $\Delta\rho$ is a factor of 5 (Rb-4) and this is accompanied with an increase in the impurity content from 5 atomic % to 40 ± 10 atomic % (50% in top layer and 30% in bottom layer).

The residual resistivity ($\rho_{40K}$) can be corrected to obtain intra-grain resistivity values using the Rowell analysis [5]. This procedure is invoked to isolate the influence of intra-grain resistivity from the inter-grain connectivity in the electrical properties. Rb-2, which shows improved inter-grain connectivity also shows the largest increase in intra-grain resistivity after thermal treatment. This could be due to increased carrier scattering by Rb or a Rb-containing phase present within the grain. Palnichenko et al. [1] have reported the presence of a phase with cubic symmetry in their Rb-doped samples. Since other samples in this study do not show a large increase in intra-grain resistivity, it is possible that in these samples, dopants (Rb/Cs) are mostly present at grain boundaries and only a small fraction of dopant is present within the grain.

*3.2.3 Effect on critical current and slope of upper critical field.* After thermal treatment, we observe a significant drop in critical current density (table 2). At this point, we are not certain if this drop due to annealing or the presence of Rb or Cs in the sample. We also note that doping and/or thermal treatment does not have a measurable effect on the slope of upper critical field ($dH_{c2}/dT$), again suggesting that the dopants have no effect on the superconducting properties of the $MgB_2$ grains.

Our results show that the Rowell analysis is useful in identifying the influence of the inter-granular connectivity and intra-granular resistivity on the electrical properties, and it also may suggest some limitations. For example, the $\rho_{40}$ can be corrected by a factor that represents the aerial current-carrying fraction ($\gamma = \rho_{40}/\rho_{40,ideal} \cong \rho_{40} / 7.4\ \mu\Omega.cm$) [6,7] to obtain a $\rho_{40K,corrected}$ that is characteristic of the intra-granular resistivity alone. Measurements of $T_c$ and the corrected value of the residual resistivity of $MgB_2$, after both ion and neutron irradiation [6,8] suggest a universal linear dependence, with $T_c$ extrapolating to 0 K for $\rho_{40k,corrected}$ values of ~90 $\mu\Omega.cm$. In the present study, however, the values of $T_c$, for the corrected values of $\rho_{40K}$ (table 2) are higher than would be expected for the universal dependence. Furthermore, the $J_c$ is found to decrease by a factor of over 25, and yet the connectivity (given by $\Delta\rho = \rho_{300}-\rho_{40}$) only changes by a factor of ~6. The accuracy of the Rowell analysis in Rb/Cs doped samples might be limited due to large deviations from the $MgB_2$ chemical composition, possibly resulting from the annealing process and potential reaction of magnesium and/or boron with other chemical species (i.e. Rb/Cs).

## 4. Conclusion

Rb and Cs with concentrations ranging from 0 to 7 atomic % have been incorporated into $MgB_2$ thin films by annealing at temperatures up to 350 °C with elemental alkali metals in an evacuated and then sealed quartz ampoule. In contrast to the reports for bulk $MgB_2$ doped with heavy alkali and alkaline earth metals, we did not find evidence for enhanced transition temperatures (> 50 K) [1] in either Rb or Cs-doped MBE grown thin films. The $T_c$ and $dH_{c2}/dT$ of thin films remain unaffected by doping. The increase in the level of oxygen and carbon contamination after the doping treatment can be related to the unusually-high reactivity of Rb and Cs. The resulting presence of impurities at the grain boundaries leads to an increase in resistivity and drop in $J_c$ of the doped films.

**Acknowledgements**

This work was supported by NSF under grant No. DMR-0514592 and ONR under contract number N00014-05-1-0105. We acknowledge use of facilities in the center for Solid State Science at ASU.